\documentclass[12pt,english]{article}\usepackage[]{graphicx}\usepackage{xcolor}
\makeatletter
\def\maxwidth{ %
  \ifdim\Gin@nat@width>\linewidth
    \linewidth
  \else
    \Gin@nat@width
  \fi
}
\makeatother

\definecolor{fgcolor}{rgb}{0.251, 0.251, 0.282}
\makeatletter
\@ifundefined{AddToHook}{}{\AddToHook{package/xcolor/after}{\definecolor{fgcolor}{rgb}{0.251, 0.251, 0.282}}}
\makeatother

\usepackage{framed}
\makeatletter
 {\par\unskip\endMakeFramed%
 \at@end@of@kframe}
\makeatother

\definecolor{shadecolor}{rgb}{.97, .97, .97}
\definecolor{messagecolor}{rgb}{0, 0, 0}
\definecolor{warningcolor}{rgb}{1, 0, 1}
\definecolor{errorcolor}{rgb}{1, 0, 0}
\makeatletter
\@ifundefined{AddToHook}{}{\AddToHook{package/xcolor/after}{
\definecolor{shadecolor}{rgb}{.97, .97, .97}
\definecolor{messagecolor}{rgb}{0, 0, 0}
\definecolor{warningcolor}{rgb}{1, 0, 1}
\definecolor{errorcolor}{rgb}{1, 0, 0}
}}
\makeatother
\newenvironment{knitrout}{}{} 

\usepackage{alltt}
\usepackage[a4paper]{geometry}
\usepackage{fullpage}
\usepackage{natbib}
\usepackage{algorithm}
\usepackage{graphicx}
\usepackage{bbm}
\usepackage{alphalph}

\usepackage[utf8]{inputenc}
\usepackage[T1]{fontenc}
\usepackage{tikz}
\usepackage{float}

\usepackage[colorlinks = true,linkcolor = blue, citecolor = blue]{hyperref}
\usepackage{booktabs}
\usepackage{amssymb,amsmath}

\def\R{\mathbb{R}}

\def\Sigmag{\mathbf{\Sigma}}

\def\Omegag{\mathbf{\Omega}} 

\def\var{{\rm var}}

\def\cov{{\rm cov}}

\def\E{{\rm E}}

\def\Vb{{\bf V}}

\def\cerc#1{\mathring{#1}}

\def\betag{\boldsymbol{\beta}} 
\def\mug{\boldsymbol{\mu}} 

\def\Sigmag{\mathbf{\Sigma}}
\def\Omegag{\mathbf{\Omega}} 

\def\Yb{{\bf Y}}
\def\Db{{\bf D}}

\def\yb{{\bf y}}
\def\zb{{\bf z}}

\def\xb{{\bf x}}

\def\Bb{{\bf B}}

\def\Nb{{\bf N}}

\def\Xb{{\bf X}}
\def\Zb{{\bf Z}}

\def\Db{{\bf D}}

\def\Wb{{\bf W}}
\def\0b{{\bf 0}}
\def\1b{{\bf 1}}

\newcommand{\myalphafoot}
{
\renewcommand{\thefootnote}{\alph{footnote}}
}

\title{An Efficient Approach for Statistical Matching of Survey Data Through Calibration, Optimal Transport and Balanced Sampling}
\myalphafoot
\author{\myalphafoot Rapha\"el Jauslin\footnotemark[1]~ and Yves Till\'e\footnotemark[1]}
\date{}
\footnotetext[1]{Institute of statistics, University of Neuchatel, Av. de Bellevaux 51, 2000 Neuchatel, Switzerland\\ (E-mail: raphael.jauslin@unine.ch)}

\usepackage{layouts}

\usetikzlibrary{patterns}
\definecolor{blue}{RGB}{39,78,135}
\definecolor{grey}{RGB}{125,125,125}
\definecolor{lgrey}{RGB}{100,100,100}
\definecolor{grey3}{RGB}{200,200,200}
\definecolor{grey2}{RGB}{166,166,166}
\definecolor{grey1}{RGB}{133,133,133}

\IfFileExists{upquote.sty}{\usepackage{upquote}}{}
\begin{document}

\maketitle

\begin{abstract}
	Statistical matching aims to integrate two statistical sources. These sources can be two samples or a sample and  the entire population.  If two samples have been selected from the same population and information has been collected on different variables of interest, then it is interesting to match the two surveys to analyse, for example, contingency tables or covariances.  In this paper, we propose an efficient method for matching two samples that may each contain a weighting scheme. The method matches the records of the two sources. Several variants are proposed in order to create a directly usable file integrating data from both information sources.
	
\end{abstract}

\maketitle

\newpage

%
%

\section{Introduction}

Integrating data from different sources represents a major challenge in statistics. \citet{yang2020statistical} and \citet{kim2020data} discussed a set of methods for data integration. Statistical matching is the field of statistics that deals with the best way to merge two different files by matching units based on a group of common variables \citep{d2006statistical,d2019statmatch}.

\citet{ren:98} distinguishes between two types of analysis. The macro approach focuses on estimating a full joint distribution between the respective variable of interest in the two samples. This can be, for example, a covariance matrix, a correlation or a contingency table. The objective of the micro approach intends to complement one file with information from the other, imputation to correct non-response is an example related to this approach \citep{haz:09,chen2019recent}. \citet{kim:ber:par:16} use the technique of fractional imputation to perform statistical matching.

In this paper, we propose an efficient method for statistical matching. Units are matched based on the proximity of a group of variables measured in both surveys. Moreover, both sources can either have common units or have an empty intersection. One of the two sources may even contain the entire population. In addition, we impose a set of constraints in order to take advantage of all the information available in the two sources. This method can also be used to realize imputations and so can also be used to micro approach analyses.

Both sources of information may contain a weighting system that allows the files to be extrapolated to the full population. These weights are usually calculated to take into account inclusion probabilities, non-response treatment and calibration. In official statistics, the calibration methods have been proposed by \citet{dev:sar:92} and \citet{dev:sar:sau:93} to adjust survey data on census data or a register. Calibration can also be used to adjust or harmonize several surveys from different sources \citep[see][and references therein]{guan:till:2017}. \citet{dud:18} through an unpublished paper, proposed to use the idea of optimal transport on internet audience data.

We have set out a series of recommendations that a matching method should follow: The method should match common units as a priority. The result of the matching must integrate information from both sources. The matching should also take into account the weighting system. After matching, the estimated totals of the variables common to both sources must be identical to the totals before matching. An optimal matching should take advantage of all the information available in both sources.

The proposed methods therefore allow the matching of two data files but also the imputation of one file on another. First, calibration theory is used to harmonise the two samples. Then a linear program is used to perform an optimal matching while taking into account the weights. This program can be written as an optimal transport problem. Finally, the values to be matched can be selected using a balanced stratified sampling technique as presented in \citet{jau:eus:til:21} and implemented in the \texttt{R} package `StratifiedSampling' \citep{StratifiedSampling}. The methods either perform matching based on a division of weights, produce a prediction, or impute a value from one source to another.

%
%

\section{Problem and notation}

Consider a population $U=\{1,\dots,k,\dots,N\}$ of size $N$ from which two random samples $S_1$ and $S_2$ of size respectively equals to $n_1$ and $n_2$ have been selected.
It is assumed that three groups of variables can be measured on the population units. The vectors of variables $\xb_k\in \R^p,k\in U$ are measured on both units selected in $S_1$ and $S_2$.  The vectors of variables $\yb_k\in \R^q,k\in U$ are measured only on the units selected in $S_1$. The vectors of variables $\zb_k\in \R^r,k\in U$ are measured only on the units selected in $S_2$. 

Generally in survey sampling, samples are designed with complex weighting systems $v_{1k},k\in S_1$ and $v_{2\ell},\ell\in S_2$. These weights can take into account the inverses of the inclusion probabilities, a possible re-weighting to compensate questionnaire non-response and a possible calibration. Frequently, a statistical match is made when one of the two file is seen as recipient file while the other one is seen as donor file. Throughout this manuscript, we will suppose, without loss of generality, that the sample $S_1$ is the recipient file while $S_2$ is the donor file. Generally we have $n_2 > n_1$, note that one of the two sample might be the whole population.

The population totals on the common auxiliary variables are equal to:
$$
\Xb = \sum_{k\in U} \xb_k.
$$
It can be estimated either by the sample $S_1$ or the sample $S_2$, which are ma\-thematically written:
\begin{equation}\label{eq:totals}
	\widehat{\Xb}_{v1} = \sum_{k\in S_1} v_{1k}\xb_k,~~ \widehat{\Xb}_{v2} = \sum_{\ell\in S_2} v_{2\ell}\xb_\ell.
\end{equation}
Following the same idea on the variables of interest, the totals
$$
\Yb = \sum_{k\in U} \yb_k \text{ and }  \Zb = \sum_{k\in U} \zb_k,
$$
can be estimated using the following estimators:
$$
\widehat{\Yb}_{v1} = \sum_{k\in S_1} v_{1k}\yb_k,~~
\widehat{\Zb}_{v2} = \sum_{\ell\in S_2} v_{2\ell}\zb_\ell.
$$

In the micro approach, the two samples  $S_1$ and $S_2$ are merged into a single usable file, while, the macro approach focuses on the joint distribution of the variables of interest. Under the usual hypothesis that, conditionally to the variables $\xb_k$, the variables $\yb_k$ and $\zb_k$ are independent, the relationships between the variables $\yb_k$ and $\zb_k$ can be analyzed.  For example, if the variables $\yb_k$ and $\zb_k$ are dummy variables with respectively $q$ and $r$ variables, we could be interested in the estimation of the contingency table
\begin{equation*}
	\Nb_{yz} = \sum_{k \in U} \yb_k\zb_k^\top.
\end{equation*}
If the variables of interest are continuous, we could be interested in computing the covariance matrix of the totals
\begin{equation*}
	\Sigmag_{yz} = \cov(\Yb,\Zb).
\end{equation*}

\section{Harmonization by calibration}\label{sec:harm}
Since we are working with two different samples, we firstly harmonize the sampling weights $v_{1k}$, $k\in S_1$ and $v_{2\ell}$, $\ell\in S_2$ in order to have same totals given in Equations \eqref{eq:totals}. That means we are looking for a new weighting system
\begin{equation}\label{eq:w}
	w_{1k}, k\in S_1 \text{ and } w_{2\ell},  \ell \in S_2,
\end{equation}
such that we have the following results:
\begin{equation*}
	\widehat{\Xb}_{w1} =\sum_{k\in S_1} w_{1k}\xb_k = \widehat{\Xb}_{w2} = \sum_{\ell\in S_2} w_{2\ell}\xb_\ell.
\end{equation*}

One aspect that must be taken into account is the intersection between $S_1$ and $S_2$.
Several cases can occur, the sample $S_1$ can be included in $S_2$ or vice versa, the intersection can also be empty. Let $n_{12}=\#(S_1\cap S_2)$ denote the size of the intersection of the two samples. \citet{guan:till:2017} analyse estimators of the form $\widehat{\Xb}_\alpha=\alpha\widehat{\Xb}_{v1}+(1-\alpha)\widehat{\Xb}_{v2}$. They showed that, when $p = 1$, to best estimate $\Xb$ using both $S_1$ and $S_2$, the value of $\alpha$ must be equal to
$$
\alpha^{\text{opt}} =
\frac{
	\var(\widehat{\Xb}_{v2})-\cov(\widehat{\Xb}_{v1},\widehat{\Xb}_{v2})
}
{
	\var(\widehat{\Xb}_{v1})+\var(\widehat{\Xb}_{v2})-\cov(\widehat{\Xb}_{v1},\widehat{\Xb}_{v2})
}.
$$
This optimal value minimizes the variance of $\widehat{\Xb}_\alpha$. However, it depends on unknown variances and on a covariance that must be estimated. Since variance estimators are particularly unstable, we may find ourselves far from the optimal estimator.  \citet{guan:till:2017} suggest to use a proxy value for $\alpha^{\text{opt}}$ that only depends on the sample sizes and the size of the overlapped sample:
\begin{equation}\label{eq:alphastar}
	\alpha^* =\frac{n_1-n_{12}}{n_1+n_2-2\; n_{12}}.
\end{equation}
One can then construct the estimator $\widehat{\Xb}^*=\alpha^*\widehat{\Xb}_{v1}+(1-\alpha^*)\widehat{\Xb}_{v2}$.
In particular, if $S_2\subset S_1$, then $\alpha^*=1$ and $\widehat{\Xb}^*=\widehat{\Xb}_{v1}. $
Moreover, if $S_1\cap S_2=\emptyset$, then $\alpha^*=n_1/(n_1+n_2).$

In order to compute the two new weighting systems \eqref{eq:w} close to $v_{1k},k\in S_1$ and $v_{2\ell},\ell\in S_2$,  the two samples are calibrated $\widehat{\Xb}^*$. If $G_k(w_{1k},v_{1k})$ is one of the pseudo-distance defined in \citet{dev:sar:92}, one can search the weighting systems that solve the following problem:
\begin{equation*}\left\{\begin{array}{ll}
		\displaystyle \text{minimize} & \displaystyle \sum_{k\in S_1} G_k(w_{1k},v_{1k}) \text{ and } \sum_{\ell\in S_2} G_\ell(w_{2\ell},v_{2\ell}) \\
		\text{subject to} & \displaystyle\sum_{k\in S_1} w_{1k}\xb_k =\sum_{\ell\in S_2} w_{2\ell}\xb_\ell = \widehat{\Xb}^* ,\\
\mbox{}
		&\displaystyle \widehat{N}^* = \sum_{k\in S_1} w_{1k} = \sum_{\ell\in S_2} w_{2\ell} =  \alpha^* \sum_{k\in S_1} v_{1k} + (1-\alpha^*)\sum_{\ell\in S_2} v_{2\ell}.
	\end{array}\right.\end{equation*}


The calibration problem must ensure that the new weights remain positive. This can be obtained, for example, by taking as pseudo-distance the divergence of Kullback-Leibler, i.e.
$G_k(w_{1k},v_{1k}) = w_{1k}\log w_{1k}/v_{1k}.$
Thus, the new weights obtained have the same sum:
$$
\sum_{k\in S_1} w_{1k} = \sum_{\ell\in S_2} w_{2\ell}.
$$
They also allow us to define new and more coherent estimators for $\Xb,\Yb$ and $\Zb$
$$
\widehat{\Xb}_{w1} =\sum_{k\in S_1} w_{1k}\xb_k = \widehat{\Xb}_{w2} = \sum_{\ell\in S_2} w_{2\ell}\xb_\ell,
$$
$$
\widehat{\Yb}_{w1} =\sum_{k\in S_1} w_{1k}\yb_k \text{ and } \widehat{\Zb}_{w2} = \sum_{\ell\in S_2} w_{2\ell}\zb_\ell.
$$

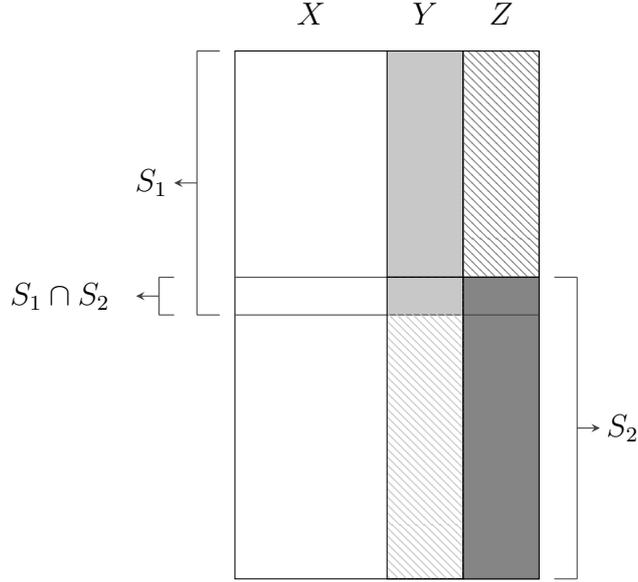
\begin{figure}[htb!]
	\begin{center}  
		\begin{tikzpicture}[scale=1]
			
			\draw[black] (-2,-3) rectangle (2,4);

			\fill[grey3] (0,0.5) rectangle (1,4);	
			\draw[black][thin] (0,1) rectangle (1,4);	
			\fill[grey1] (1,-3) rectangle (2,1);	
			\draw[black][thin] (1,-3) rectangle (2,1);

			\draw[darkgray][thin] (-2,1.0) -- (2.0,1.0);
			\draw[darkgray][thin] (-2,0.5) -- (2.0,0.5);
			
			\node at (-3.1,2.25) {$S_1$};					
			\draw[darkgray][thin] (-2.2,0.5) -- (-2.5,0.5);
			\draw[darkgray][thin] (-2.2,4.0) -- (-2.5,4.0);				
			\draw[darkgray][thin] (-2.5,4.0) -- (-2.5,0.5);					
			\draw[->,>=stealth][darkgray][thin] (-2.5,2.25) -- (-2.8,2.25);

			\draw[darkgray][thin] (-3.0,1.0) -- (-2.8,1.0);
			\draw[darkgray][thin] (-3.0,0.5) -- (-2.8,0.5);
			\draw[darkgray][thin] (-3.0,0.5) -- (-3.0,1.0);
			\draw[->,>=stealth][darkgray][thin] (-3.0,0.75) -- (-3.3,0.75);
			\node at (-4.3,0.75) {$S_1\cap S_2$};
			
			\node at (3.1,-1) {$S_2$};
			
			\draw[darkgray][thin] (2.2,1.0) -- (2.5,1.0);
			\draw[darkgray][thin] (2.2,-3.0) -- (2.5,-3.0);				
			\draw[darkgray][thin] (2.5,-3.0) -- (2.5,1.0);					
			\draw[->,>=stealth][darkgray][thin] (2.5,-1) -- (2.8,-1);			
			
			\draw[black,pattern=north west lines,pattern color=grey1][thin] (1,1) rectangle (2,4);
			\draw[black,pattern=north west lines,pattern color=grey3][thin] (0,-3) rectangle (1,1);
			
			\node at (-1,4.5) {$X$};
			\node at (0.5,4.5) {$Y$};
			\node at (1.5,4.5) {$Z$};

		\end{tikzpicture}
	\end{center} \caption{Representation of the statistical matching with intersection. Hatched area are unknown quantities in each samples.}
	\label{fig:4}
\end{figure}

%
%
\section{Renssen's methods}
A method for estimating contingency table is developed in \citet{ren:98} and more recently presented in \citet{d2006statistical}. The general idea consists of harmonizing the weighting systems as explained in the Section \ref{sec:harm} and then uses the matching variables $\xb_k$ to create linear models to get an estimated contingency table. At the first step, regression coefficients $\betag_{yx}$ and $\betag_{zx}$  are computed  from the samples $S_1$ (respectively $S_2$) by using the weights, $w_{1k}$, $k\in S_1$ (respectively $w_{2\ell}$, $\ell\in S_2$). Using a weighted linear model, the following coefficients are obtained:
$$
	\widehat{\betag}_{yx} = \left(\sum_{k \in S_1} w_{1k}\yb_k\xb_k^\top \right)\left(\sum_{k \in S_1} w_{1k}\xb_k\xb_k^\top \right)^{-1},
$$
$$
	\widehat{\betag}_{zx} = \left(\sum_{\ell \in S_2} w_{2\ell}\zb_\ell\xb_\ell^\top \right)\left(\sum_{\ell \in S_2} w_{2\ell}\xb_\ell\xb_\ell^\top \right)^{-1}.
$$

The contingency table is then estimated using the matrix product:
\begin{equation*}
	\widehat{\Nb}_{yz}^{ren} = \widehat{\betag}_{yx}\left(\alpha^* \sum_{k \in S_1} w_{1k}\xb_k\xb_k^\top + (1 - \alpha^*)\sum_{\ell\in S_2} w_{2\ell}\xb_\ell\xb_\ell^\top \right) \widehat{\betag}_{zx}^\top,
\end{equation*}
where $\alpha^*$ is the coefficient \eqref{eq:alphastar} that depends on the value $n_{12}$. Renssen's method can be easily generalized to continuous case, but some assumptions must be satisfied on  variables $\yb_k$, $k\in S_1$ and $\zb_\ell$, $\ell\in S_2$. For more information, we refer the reader to the article of \citet{ren:98} and the book written by \citet{d2006statistical}.

%
%
\section{Matching by optimal transport}\label{sec:opt}

The main idea of our method uses the optimal transport to perform a statistical matching. Optimal transport is an old mathematical problem that consists of finding the best solution to minimize the cost of transporting some quantities of goods from a given set of locations to a given set of destinations. In its simple case, the optimal transport problem can be solved with a linear program. However, it has been a very fruitful topic in statistics for the past 10 years and it is strongly related to the notion of Wasserstein distance. We refer the reader to \citet{pan:zem:2020} for more information on optimal transport and the Wasserstein distance.

In the particular case of statistical matching, we use the optimal transport problem to match the units of two different samples. By reminding that $S_1$ denotes the recipient sample while $S_2$ stands for the donor sample, the idea is then to see which units of $S_2$ can be associated to a particular unit $k \in S_1$. We start by computing a $n_1 \times n_2$ matrix $\Db$ containing the distances between the units of $S_1$ and the units of $S_2$. We can for example use the usual Euclidean distance or a Mahalanobis distance defined as follows:
$$
d^2(k,\ell) = (\xb_k-\xb_\ell)^\top\widehat{\Omegag}_{xx}^{-1} (\xb_k-\xb_\ell),
$$
where
$$
\widehat{\Omegag}_{xx} = \frac{1}{\widehat{N}^*}
\left\{ \alpha^*\sum_{k\in S_1} w_{1k}(\xb_k - \widehat{\overline{\Xb}})(\xb_k-\widehat{\overline{\Xb}})^\top
+  (1-\alpha^*)\sum_{\ell\in S_2} w_{2\ell}(\xb_\ell - \widehat{\overline{\Xb}})(\xb_\ell-\widehat{\overline{\Xb}})^\top \right\},
$$
and
$$
\widehat{\overline{\Xb}}= \frac{\widehat{\Xb}^*}{\widehat{N}^*}.
$$
Then, we  search for weights $W_{k\ell}$ for each couple $k\in S_1,\ell\in S_2.$ To do this, we solve the following linear program:

\begin{equation*}\left\{\begin{array}{rl}
		\displaystyle \text{minimize} & \displaystyle \sum_{k\in S_1}\sum_{\ell\in S_2} W_{k\ell} \; d(k,\ell) \\
		\text{subject to} & \displaystyle\sum_{k\in S_1} W_{k\ell} = w_{2\ell} \text{ for all } \ell \in S_2, \\
		&\displaystyle \sum_{\ell\in S_2} W_{k\ell} = w_{1k} \text{ for all } k \in S_1,\\
		& \displaystyle W_{k\ell}\ge 0, \text{ for all } k \in S_1,\text{ and }\ell \in S_2,
	\end{array}\right.\end{equation*}
where $W_{kk} = \min( w_{1k}, w_{2k}),$ for all couples of identical units in $S_1$  and $S_2$. These constraints force the matching of identical units that can be selected from both samples. This linear program is nothing more than an optimal transport problem for which it exists many efficient implementations.

Most of the $W_{k\ell}$ weights are zero. It is therefore not necessary to manipulate a large matrix of data. The realized calibration is not adversely affected in the linear program. Thus we have
$$
\sum_{k\in S_1} \sum_{\ell\in S_2} W_{k\ell} \xb_k =\sum_{k\in S_1} \sum_{\ell\in S_2} W_{k\ell} \xb_\ell =  \widehat{\Xb}^*.
$$

The output of the linear program ends with a matrix of weights $\Wb$ of size $(n_1 \times n_2)$. The non-zero entries in the $i$th rows of the matrix $\Wb$ contain the corresponding weights of the matched units in the sample $S_2$. We generally do not have a one-to-one match, which means that for each unit $k$ in $S_1$ we have more than one unit with weights not equal to 0 in $S_2$. The next two sections propose two different ways to obtain, from the output of the optimal transport, a file where each unit from $S_1$ has only one imputed unit from $S_2$. Without loss of generality, in the following development, we suppose that sample $S_1$ is completed by realizing a prediction from $S_2$.

	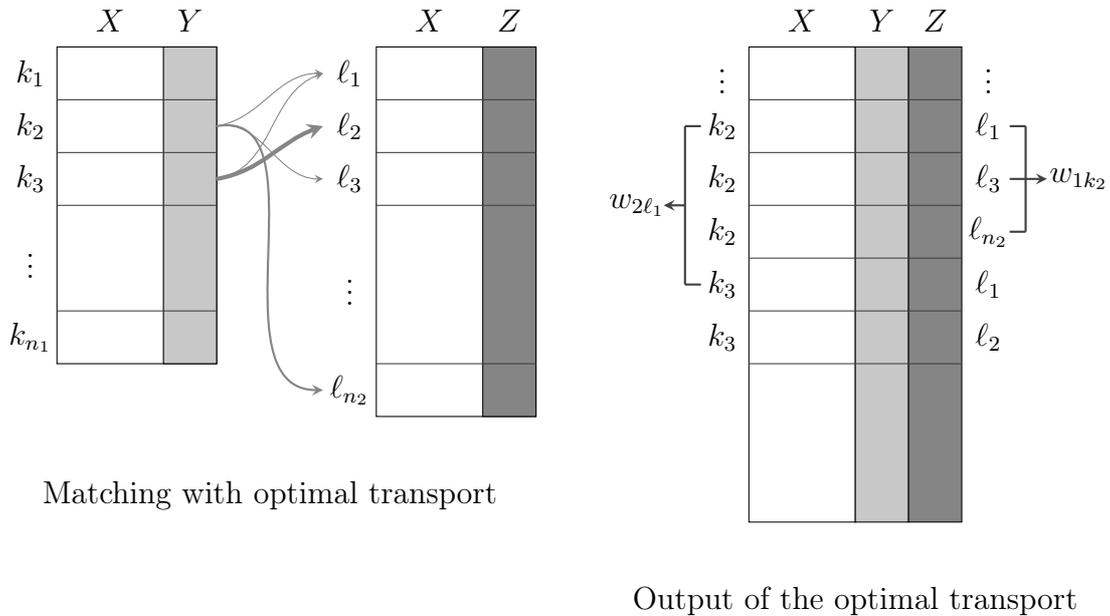
\begin{figure}[htb!]
	\begin{center}  
		\begin{tikzpicture}[scale=0.7]
			
			\draw[black] (-4,-2) rectangle (-1,4);				
			\draw[black] (2,-3) rectangle (5,4);

			\fill[grey3] (-2,-2) rectangle (-1,4) ;
			\draw[black] (-2,-2) rectangle (-1,4) ;
			
			\fill[grey1] (4,-3) rectangle (5,4) ;
			\draw[black] (4,-3) rectangle (5,4) ;				
			
			\draw[darkgray][thin] (-4,3.0) -- (-1,3.0);
			\draw[darkgray][thin] (-4,2.0) -- (-1,2.0);
			\draw[darkgray][thin] (-4,1.0) -- (-1,1.0);
			\draw[darkgray][thin] (-4,-1.0) -- (-1,-1.0);
			
			\draw[darkgray][thin] (2,3.0) -- (5,3.0);
			\draw[darkgray][thin] (2,2.0) -- (5,2.0);
			\draw[darkgray][thin] (2,1.0) -- (5,1.0);
			\draw[darkgray][thin] (2,-2.0) -- (5,-2.0);
			
			\node at (-3,4.5) {$X$};
			\node at (-1.5,4.5) {$Y$};
			
			\node at (3,4.5) {$X$};
			\node at (4.5,4.5) {$Z$};

			\draw[grey1][->,>=stealth,thick] (-1,2.5) to [out=10,in = 180] (1,-2.5);
			\draw[grey1][->,>=stealth,thin] (-1,2.5) to [out=10,in = 180] (1,1.5);
			\draw[grey1][->,>=stealth] (-1,2.5) to [out=10,in = 180] (1,3.5);
			
			\draw[grey1][->,>=stealth,ultra thick] (-1,1.5) to [out=10,in = 200] (1,2.5);
			\draw[grey1][->,>=stealth] (-1,1.5) to [out=10,in = 200] (1,3.5);
			
			\node at (-4.5,3.5) {$k_1$};
			\node at (-4.5,2.5) {$k_2$};
			\node at (-4.5,1.5) {$k_3$};				
			\node at (-4.5,-0) {$\vdots$};

			\node at (-4.5,-1.5) {$k_{n_1}$};	
			\node at (1.5,3.5) {$\ell_1$};
			\node at (1.5,2.5) {$\ell_2$};
			\node at (1.5,1.5) {$\ell_3$};
			\node at (1.5,-0.5) {$\vdots$};				
			\node at (1.5,-2.5) {$\ell_{n_2}$};

			\node at (-0,-4.5) {Matching with optimal transport};
			\node at (11,-6.5) {Output of the optimal transport};

			
			\draw[black] (9,-5) rectangle (11,4);
			
			\fill[grey3] (11,-5) rectangle (12,4) ;				
			\draw[black] (11,-5) rectangle (12,4);

			\fill[grey1] (12,-5) rectangle (13,4) ;
			\draw[black] (12,-5) rectangle (13,4);

			\draw[darkgray][thin] (9,3.0) -- (13,3.0);
			\draw[darkgray][thin] (9,2.0) -- (13,2.0);
			\draw[darkgray][thin] (9,1.0) -- (13,1.0);
			\draw[darkgray][thin] (9,0.0) -- (13,0.0);
			\draw[darkgray][thin] (9,-1.0) -- (13,-1.0);
			\draw[darkgray][thin] (9,-2.0) -- (13,-2.0);

			\node at (10,4.5) {$X$};
			\node at (11.5,4.5) {$Y$};
			\node at (12.5,4.5) {$Z$};

			\node at (8.5,3.5) {$\vdots$};
			\node at (8.5,2.5) {$k_2$};
			\node at (8.5,1.5) {$k_2$};
			\node at (8.5,0.5) {$k_2$};
			\node at (8.5,-0.5) {$k_3$};
			\node at (8.5,-1.5) {$k_3$};
			
			\draw[darkgray][thick] (8.1,2.5) -- (7.8,2.5);
			\draw[darkgray][thick] (8.1,-0.5) -- (7.8,-0.5);
			\draw[darkgray][thick] (7.8,-0.5) -- (7.8,2.5);
			\draw[darkgray][->,>=stealth,thick] (7.8,1) -- (7.4,1);
			\node at (6.9,1) {$w_{2\ell_1}$};

			\node at (13.5,3.5) {$\vdots$};
			\node at (13.5,2.5) {$\ell_1$};
			\node at (13.5,1.5) {$\ell_3$};
			\node at (13.5,0.5) {$\ell_{n_2}$};	
			\node at (13.5,-0.5) {$\ell_1$};	
			\node at (13.5,-1.5) {$\ell_2$};

			\draw[darkgray][thick] (13.9,2.5) -- (14.2,2.5);
			\draw[darkgray][thick] (13.9,1.5) -- (14.2,1.5);
			\draw[darkgray][thick] (13.9,0.5) -- (14.2,0.5);
			\draw[darkgray][thick] (14.2,2.5) -- (14.2,0.5);
			\draw[darkgray][->,>=stealth,thick] (14.2,1.5) -- (14.6,1.5);
			\node at (15.2,1.5) {$w_{1k_2}$};

		\end{tikzpicture}
	\end{center} \caption{Insight of the statistical matching using optimal transport.  }
	\label{fig:example}
\end{figure}

%
%

\subsection{Matching by using prediction}
We can do a prediction by computing the weighted averages of the $\xb_\ell$ and $\zb_\ell$ of $S_2$. Formally, this gives the following quantity to compute:
$$
q_{k\ell} =\frac{W_{k\ell}}{\sum_{\ell\in S_2} W_{k\ell}}= \frac{W_{k\ell}}{w_{1k}}, \text{ for all } k\in S_1, \ell \in S_2.
$$
By using these new weights, we can then compute a prediction of the $\xb_k$ and the $\zb_k$ $k\in S_1$,
$$
\widehat{\xb}_k =  \sum_{\ell \in S_2} q_{k\ell} \xb_\ell \text{ and }
\widehat{\zb}_k =  \sum_{\ell \in S_2} q_{k\ell} \zb_\ell, \text{ for all } k\in S_1.
$$
The matching quality can be evaluated by comparing the $\xb_k$ with the  predictions $\widehat{\xb}_k$.
For the predicted values $\widehat{\xb}_k$, the calibration is always valid. Indeed that
$$
\sum_{k \in S_1} w_{1k} \widehat{\xb}_k = \sum_{k \in S_1} w_{1k} \xb_k =\widehat{\Xb}^*.
$$
However, the interest of the procedure is that we now have predicted values $\widehat{\zb}_k$ for each unit of $S_1$ whereas these variables were only measured on $S_2$.

%
%

\subsection{Matching by using stratified balanced sampling}

As explained in Section \ref{sec:opt}, the output of the optimal transport has generally repetition of some units. In some cases, we are interested in obtaining a synthetic file without any repetition of some units. In this section, we propose an imputation method based on the optimal transport result. We propose to use a balanced sampling method to select from the repetition stratum of a unit $k$ of $S_1$, only one unit of the units of $S_2$. This is in line with hypothesis that $S_2$ is the donor file while the $S_1$ is the recipient file. To do this, we randomly generate a matrix of Bernoulli random variable  $a_{k\ell},k\in S_1,\ell \in S_2,$ where $a_{k\ell}$ is 1 if
unit $\ell\in S_2$ is imputed to unit $k\in S_1.$ Since each unit $k$ of $S_1$ can only receive one imputation, we must have
$$
\sum_{\ell\in S_2} a_{k\ell} = 1, \text{ for all } k\in S_1.
$$

We now want to generate the random matrix of $a_{k\ell}$ with expectations $\E(a_{k\ell}) = q_{k\ell}$ in such a way that the following system of equations is satisfied at best
$$
\sum_{k\in S_1} \sum_{\ell\in S_2} \frac{a_{k\ell}W_{k\ell}}{q_{k\ell}} \xb_\ell \approx \sum_{k\in S_1} \sum_{\ell\in S_2} W_{k\ell} \xb_\ell = \widehat{\Xb}^*,
$$
$$
\sum_{k\in S_1} \sum_{\ell\in S_2} \frac{a_{k\ell}W_{k\ell}}{q_{k\ell}} \zb_\ell \approx \sum_{k\in S_1} \sum_{\ell\in S_2} W_{k\ell} \zb_\ell
$$
and with
$$
\sum_{\ell\in S_2} a_{k\ell} = 1, \text{ for all } k\in S_1.
$$
This sampling problem is known as `stratified balanced sampling' \citep[see][]{jau:eus:til:21,has:til:14,StratifiedSampling}.
Indeed, each unit $k$ of $S_1$ can be seen as a stratum for which a unit $\ell$ of $S_2$ must be selected.

The imputed values are then
$$
\cerc{\xb}_k =  \sum_{\ell \in S_2} a_{k\ell} \xb_\ell \text{ and }
\cerc{\zb}_k =  \sum_{\ell \in S_2} a_{k\ell} \zb_\ell, \text{ for all } k\in S_1.
$$
Again, we have
$$
\sum_{k \in S_1} w_{1k} \cerc{\xb}_k \approx \sum_{k \in S_1} w_{1k} \xb_k =\widehat{\Xb}^*.
$$
However, the interest of the procedure is that we now have values $\cerc{\zb}_k$ for each unit of $S_1$ whereas these variables were only measured on $S_2$.

If $\E_q(.)$ is the expectation to the $a_{k\ell}$ conditionally to $S_1$ and $S_2$, then, for all $k\in S_1,$ we have,
$$
	\E_q(\cerc{\xb}_k) =  \sum_{\ell \in S_2} \E_q(a_{k\ell}) \xb_\ell=  \sum_{\ell \in S_2} q_{k\ell} \xb_\ell =\widehat{\xb}_k,
$$
and
$$
	\E_q(\cerc{\zb}_k) =  \sum_{\ell \in S_2} \E_q(a_{k\ell}) \zb_\ell=  \sum_{\ell \in S_2} q_{k\ell} \zb_\ell =\widehat{\zb}_k.
$$

%
%

\section{Analysis of the data}

Once the optimal transport is performed, different analysis are possible. We can either work directly with the optimal transport result or use the prediction or the imputations methods.
This gives us five possibilities to analyse the data.

\begin{enumerate}
	\item Use full results of the optimal transport problem $(\xb_k,\xb_\ell,\yb_k,\zb_\ell,W_{k\ell},k\in S_1,\ell \in S_2)$.
	\item Use the predicted values $(\xb_k,\widehat{\xb}_k,\yb_k,\widehat{\zb}_k,w_{1k},k\in S_1)$ by predicting the values of $k\in S_1$.
	\item Use the imputed values $(\xb_k,\cerc{\xb}_k,\yb_k,\cerc{\zb}_k,w_{1k},k\in S_1)$ by imputing the values of $k\in S_1$.
	\item Use the predicted values $(\xb_\ell,\widehat{\xb}_\ell,\widehat{\yb}_\ell,\zb_\ell,w_{2\ell},\ell\in S_2)$ by predicting the values of $\ell\in S_2$.
	\item Use the imputed values $(\xb_\ell,\cerc{\xb}_\ell,\cerc{\yb}_\ell,\zb_\ell,w_{2\ell},\ell\in S_2)$ by imputing the values of $\ell\in S_2$.
\end{enumerate}

For all these possibilities, the estimations of the means are completely consistent for the five possibilities. Indeed, we obtain
$$
	\widehat{\overline{\Zb}} =\frac{ \sum_{k\in S_1}\sum_{\ell \in S_2} W_{k\ell}  \zb_\ell }{ \sum_{k\in S_1}\sum_{\ell \in S_2} W_{k\ell} }
	= \frac{ \sum_{\ell\in S_2} w_{2\ell} \zb_\ell }{ \sum_{\ell\in S_2} w_{2\ell}} =  \frac{ \sum_{k\in S_1} w_{1k} \widehat{\zb}_k }{ \sum_{k\in S_1} w_{1k}  }
	\approx \frac{ \sum_{k\in S_1} w_{1k}  \cerc{\zb}_k }{ \sum_{k\in S_1} w_{1k} },
$$
and
$$	\widehat{\overline{\Yb}} =\frac{ \sum_{k\in S_1}\sum_{\ell \in S_2} W_{k\ell}  \yb_k }{ \sum_{k\in S_1}\sum_{\ell \in S_2} W_{k\ell} }
	= \frac{ \sum_{k\in S_1} w_{1k} \yb_k }{ \sum_{k\in S_1} w_{1k}} ~=~ \frac{ \sum_{\ell\in S_2} w_{2\ell} \widehat{\yb}_\ell }{ \sum_{\ell\in S_2} w_{2\ell}}
	\approx  \frac{ \sum_{\ell\in S_2} w_{2\ell} \cerc{\yb}_\ell }{ \sum_{\ell\in S_2} w_{2\ell}}.
$$
If the variables are categorical, we can then estimate a contingency table using the results of the optimal transport matching, or the prediction on $S_1$ (respectively on $S_2$),
$$
	\widehat{\Nb}_{yz} =\sum_{k\in S_1} \sum_{\ell\in S_2} W_{k\ell} \yb_k\zb_\ell^\top
	= \sum_{k\in S_1} w_{1k} \yb_k\widehat{\zb}_k^\top =  \sum_{\ell\in S_2} w_{2\ell} \widehat{\yb}_\ell\zb_\ell^\top.
$$
We can also use the imputed values that give slightly different results,
$$ \widehat{\Nb}_{yz}^1 =  \sum_{k\in S_1} w_{1k} \yb_k\cerc{\zb}_k^\top \text{ and } \widehat{\Nb}_{yz}^2 =  \sum_{\ell\in S_2} w_{2\ell} \cerc{\yb}_\ell\zb_\ell^\top.$$

If the variables are continuous, we can estimate the covariances between the $\yb_k$ and the $\zb_\ell$ variables. We can also work indifferently from $S_1\times S_2$, $S_1$ or $S_2$. Indeed, we have
\begin{equation*}
	\begin{split}
		\widehat{\Sigmag}_{yz}
		&=\frac{1}{\widehat{N}^*} \sum_{k\in S_1}\sum_{\ell \in S_2} W_{k\ell}  (\yb_k-\widehat{\overline{\Yb}})(\zb_\ell-\widehat{\overline{\Zb}})^\top \\
		&=\frac{1}{\widehat{N}^*} \sum_{k\in S_1} w_{1k}  (\yb_k-\widehat{\overline{\Yb}})(\widehat{\zb}_k-\widehat{\overline{\Zb}})^\top \\
		&=\frac{1}{\widehat{N}^*} \sum_{\ell \in S_2} w_{2\ell}  (\widehat{\yb}_\ell-\widehat{\overline{\Yb}})(\zb_\ell-\widehat{\overline{\Zb}})^\top.
	\end{split}
\end{equation*}
As previously seen, it is also possible to use the imputed values that give slightly different results
\begin{equation*}
	\widehat{\Sigmag}_{yz}^{1}=\frac{1}{\widehat{N}^*} \sum_{k\in S_1} w_{1k}  (\yb_k-\widehat{\overline{\Yb}})(\cerc{\zb}_k-\widehat{\overline{\Zb}})^\top
\end{equation*}
and
\begin{equation*}
	\widehat{\Sigmag}_{yz}^{2}=\frac{1}{\widehat{N}^*} \sum_{\ell \in S_2} w_{2\ell}  (\cerc{\yb}_\ell-\widehat{\overline{\Yb}})(\zb_\ell-\widehat{\overline{\Zb}})^\top.
\end{equation*}
Since $\E_q(\cerc{\yb}_k) =\widehat{\yb}_k$ and $\E_q(\cerc{\zb}_k) =\widehat{\zb}_k$, then $
\E_q(\widehat{\Sigmag}_{yz}^{1}) = \E_q(\widehat{\Sigmag}_{yz}^{2}) = \widehat{\Sigmag}_{yz}.
$
The three estimators are thus very close to each other. One can thus use in an undifferentiated way $S_1\times S_2$, $S_1$ or $S_2$.

%
%

\section{Simulations}\label{sec:simu}

In this section we propose two simulations, the first one on a simulated normal dataset where the conditional independence assumption holds and a second one on the Austrian data EU-SILC (European Union Statistics on Income and Living Conditions).
\subsection{Gaussian example}
In this section we discuss a generated dataset such that the CIA is satisfied. Let a population $U$ of 10 000 units generated such that the matching variables $\xb_k\in\R^p, k\in U$, the variables recorded only in $S_1$ $\yb_k\in\R^q, k \in U$ and the variables recorded only in $S_2$, $\zb_k\in\R^r, k\in U$ are normally distributed with joint distribution equals to  

$$ f(\xb,\yb,\zb| \mug, \Sigmag) = \frac{1}{\sqrt{2\pi^{p+q+r} |\Sigmag| }}\exp{\left[-\frac{1}{2}\left\{ (\xb,\yb,\zb)^\top - \mug \right\}^\top \Sigmag^{-1} \left\{(\xb,\yb,\zb)^\top - \mug \right\}\right] },$$
where the parameters are defined as follow
$$\mug = \left(\begin{matrix}
	\mug_x\\\mug_y \\ \mug_z
\end{matrix} \right),~~~~~~\Sigmag= \left(
\begin{matrix}
	\Sigmag_{xx} & \Sigmag_{xy} & \Sigmag_{xz}\\
	\Sigmag_{yx} & \Sigmag_{yy} & \Sigmag_{yz}\\
	\Sigmag_{zx} & \Sigmag_{zy} & \Sigmag_{zz}\\
\end{matrix}
\right).
$$

To ensure that the CIA holds, the quantity $\Sigmag_{yz}$ must be determined by the following equality \citep{d2006statistical}:
$$ \Sigmag_{yz} = \Sigmag_{yx}\Sigmag_{xx}^{-1}\Sigmag_{xz}$$ 
Let suppose that $p = 3$, $q = 2$ and $r = 2$. We simulate 10~000 samples of recipient sample $S_1$ and donor $S_2$ by using simple random sampling without replacement with sizes fixed to $n_1 = 600$ and $n_2 = 3000$. The parameters of the matching variables are equal to

$$\mug_x = \left(\begin{matrix}
	0\\0 \\0
\end{matrix} \right),~~~~~~\Sigmag_{xx} = \left(
\begin{matrix}
	7.364 & 2.579 & -0.475\\ 
	2.579 & 5.694 & -0.021\\ 
	-0.475 & -0.021 & 7.864 
\end{matrix}
\right).
$$
By using the properties of the Gaussian distribution, if $\yb_k = (y_{k1} ~~ y_{k2})^\top\in \R^2, k\in U$ and $\zb_k = (z_{k1} ~~ z_{k2})^\top\in \R^2, k\in U$ are linear combination of the matching variables, the CIA holds and we can estimate the covariance matrix $\Sigmag_{yz}$. Linear combinations are then equal to
$$\begin{array}{lll}
	 y_{k1} &=& 0.2x_{k1} -0.3x_{k2} + x_{k3} + \varepsilon_k,\\ 
	 y_{k2} &=& 1.2x_{k1} + 0.4x_{k2} - 0.5x_{k3} + \varepsilon_k,\\
	 z_{k1} &=& -0.4x_{k1} + x_{k2} - 0.3x_{k3}+ \varepsilon_k,\\ 
	 z_{k2} &=& -1.4x_{k1} + 0.3x_{k2} - 0.6x_{k3}+ \varepsilon_k,\\ 
\end{array}$$
where $\varepsilon_k\sim \mathcal{N}(0,1)$ and the variance-covariance matrix is given by

$$\Sigmag_{yz} = \left(
\begin{matrix}
	-3.667 & -5.368\\
	2.627 & -9.772	
\end{matrix}
\right).
$$

\begin{table}[!h]

	\caption{\label{tab:simu_gauss}Mean squared errors and its bias-variance decomposition on 10~000 simulations of the estimation of the variance-covariance matrix $\Sigmag_{yz}$. Matrix notation $\Bb$ stands for the squared bias while $\Vb$ denote the variance of the estimator $\widehat{\Sigmag}_{yz}$.}
	\centering
	\begin{tabular}[t]{lcc|c}
	\toprule
	\addlinespace[2ex]
	\multicolumn{4}{l}{Optimal transport}\\
	\addlinespace[1ex]
	& $  \Bb_{yz}^{opt} =  \begin{pmatrix}  0.014 & 0.022 \\ 0.013 &  0.046 \end{pmatrix}$ & $ \Vb_{yz}^{opt} = \begin{pmatrix} 0.032 & 0.09 \\ 0.042 & 0.162 \end{pmatrix}$  & $ \text{MSE}(\widehat{\Sigmag}_{yz}^{opt}) = \begin{pmatrix} 0.046 & 0.113 \\ 0.056 & 0.208 \end{pmatrix}$ \\	
	\addlinespace[2ex]
	\multicolumn{4}{l}{Renssen}\\
	\addlinespace[1ex]
	& $\Bb_{yz}^{ren} =\begin{pmatrix}  0 & 0 \\ 0.011 &  0 \end{pmatrix}$ & $ \Vb_{yz}^{ren} =\begin{pmatrix} 0.104 & 0.255 \\ 0.162 & 0.568 \end{pmatrix}$  & $ \text{MSE}(\widehat{\Sigmag}_{yz}^{ren}) = \begin{pmatrix} 0.104 & 0.255 \\ 0.173 & 0.568 \end{pmatrix}$ \\
	\addlinespace[2ex]
	\multicolumn{4}{l}{Balanced imputation}\\
	\addlinespace[1ex]
	& $ \Bb_{yz}^{bal} =\begin{pmatrix}  0.011 & 0.021 \\ 0.013 &  0.045 \end{pmatrix}$ & $  \Vb_{yz}^{bal} =\begin{pmatrix} 0.055 & 0.127 \\ 0.086 & 0.235 \end{pmatrix}$  & $ \text{MSE}(\widehat{\Sigmag}_{yz}^{bal}) = \begin{pmatrix} 0.066 & 0.148 \\ 0.099 & 0.279 \end{pmatrix}$ \\
	\bottomrule
	\end{tabular}
	\end{table}

Table \ref{tab:simu_gauss} shows the mean squared errors of the simulations of the three different methods. The mean squared error result shows that optimal transport procedure and balanced imputation method are better. It is important to note here that the optimal transport is slightly biased. Optimal transport and balanced imputation do not use a model prediction to create the match. In this particular case, where everything is linear, the Renssen's method will be very powerful in creating a statistical match. Nevertheless, the mean squared error is better than using the optimal transport methods. In general, with a real dataset, the CIA does not hold, and furthermore, it is even not testable. The next section presents an example where the CIA does not hold and the optimal transport methods are better.

\subsection{EU-SILC example}

\begin{figure}[ht!]
 	\centering
 	\input{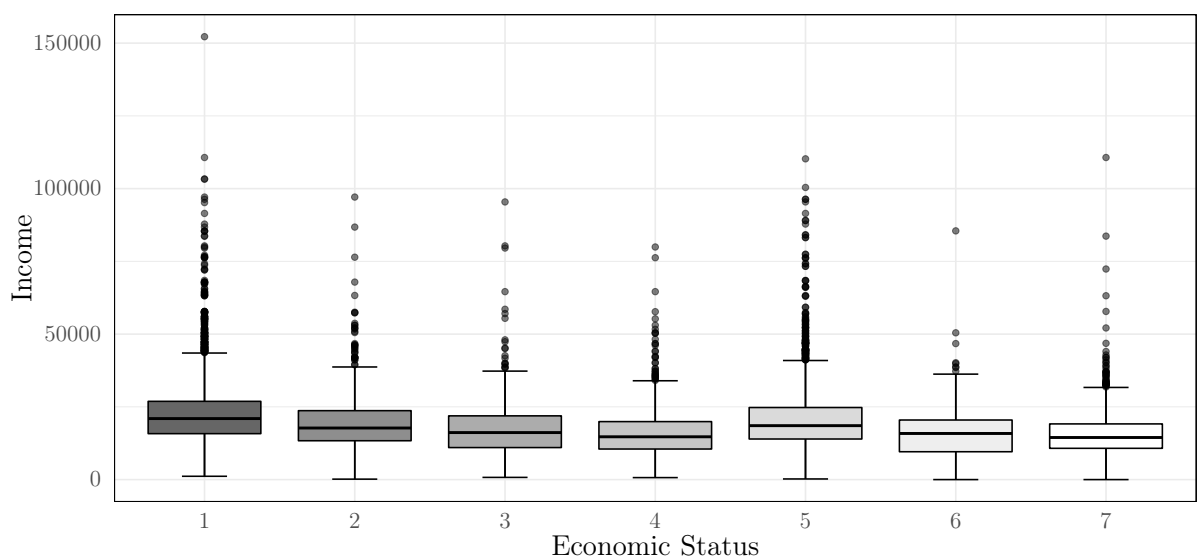}
 	\caption{Boxplot of the household income by economic status.}
 	\label{fig:boxplot}
\end{figure}

This section proposes a simulation study to see how the proposed method performs compared to the method proposed by \citet{ren:98} on the dataset \texttt{eusilc} available in the R package \citet{cran:laeken}.
This dataset contains 14~827 observations and 28 variables. It is based on real Austrian data EU-SILC (European Union Statistics on Income and Living Conditions). We slightly modified the dataset to remove the missing values. It represents then a dataset of 12~107 observations. Table \ref{tab:eusilc} shows a summary of the different variables used for the simulations. In particular, \texttt{pl030} is the categorical variable representing the economic status while \texttt{eqIncome} represents the continuous variable of household income. \texttt{pl030} is the variable of interest recorded only in $S_1$ while \texttt{eqIncome} is recorded only in $S_2$. Figure \ref{fig:boxplot} shows the income household by economic status, each simulation will estimate the average income by category.



We run simulations using stratified balanced sampling design \citep{jau:eus:til:21} with sample size $n_1 = 1000 $ for each sample $S_1$ and $n_2 = 4000$ for each sample $S_2$. Inclusion probabilities are selected such that the design respects an optimal stratification, i.e., the number of unit selected in each stratum is proportional to the product of the stratum size and the standard error of \texttt{eqIncome}. In addition, the sample are balanced on the totals of the matching variables. 

To measure the effectiveness of our proposed estimators, we use the mean squared error (MSE). Table~\ref{tab:YZ2} shows the mean squared errors of the estimation of the average income within each category based on 10~000 simulations. The table displays also the bias-variance decomposition. We can observe that the matching based on optimal transport is the best in terms of mean squared error. 


\begin{table}[htb!]
	\begin{center}
		\caption{Selected variables of the \texttt{eusilc} dataset of the \texttt{R} package developed by \citet{cran:laeken}. The first five variables are the ones used for the matching while the two last ones are the variables of interest.\label{tab:eusilc}}
		\begin{tabular}{ll}
			\addlinespace[0.5ex]
			\toprule
			\addlinespace[1ex]
			\multicolumn{2}{l}{\textbf{Matching variables}}\\
			\addlinespace[1ex]
			\texttt{hsize} & The number of persons in the household.\\
			\texttt{db040} & The federal state in which the household is located.\\
			\texttt{age} & The person’s age.\\
			\texttt{rb090} & The person’s gender. (male or female) \\
			\texttt{pb220a} & The person’s citizenship (AT, EU and Other).\\
			\addlinespace[2ex]
			\multicolumn{2}{l}{\textbf{Variables of interests}}\\
			\addlinespace[1ex]			
			\texttt{pl030} &  \begin{tabular}{lll} 1 : & working full time. \\ 2 : & working part time. \\ 3 : & unemployed. \\ 4 : & pupil, student, further training,  \\ & unpaid work experience,\\ & in compulsory military or community service. \\ 5 : & in retirement or early retirement or has given up business. \\ 6 : & permanently disabled or/and \\ & unfit to work or other inactive person. \\ 7 : & fulfilling domestic tasks and care responsibilities. \end{tabular} \\
			\addlinespace[1ex]
			\texttt{eqIncome} &  Slightly simplified version of the household income.\\
			\bottomrule
			\addlinespace[0.5ex]
		\end{tabular}
	\end{center}
\end{table}

\begin{knitrout}
\definecolor{shadecolor}{rgb}{0.973, 0.973, 0.973}\color{fgcolor}\begin{table}[!h]

\caption{\label{tab:YZ2}Results of 10~000 simulation for the estimation of average income per category. Relative root mean squared errors as well as bias-variance decomposition are calculated for each economical status. The overall mean squared error are equal to 48.304  for the optimal transport, 62.792  for the balanced imputation and  53.176 for the method of Renssen.}
\centering
\begin{tabular}[t]{llllllll}
\toprule
\multicolumn{1}{c}{ } & \multicolumn{7}{c}{Economical Status } \\
\cmidrule(l{3pt}r{3pt}){2-8}
Method & 1 & 2 & 3 & 4 & 5 & 6 & 7\\
\midrule
\addlinespace[1ex]
\multicolumn{8}{l}{MSE}\\
\hspace{1em}$opt$ & 0.072 & 0.058 & 0.158 & 0.151 & 0.02 & 0.247 & 0.241\\
\hspace{1em}$bal$ & 0.074 & 0.078 & 0.178 & 0.168 & 0.035 & 0.304 & 0.25\\
\hspace{1em}$ren$ & 0.089 & 0.027 & 0.15 & 0.227 & 0.014 & 0.23 & 0.238\\
\addlinespace[1ex]
\multicolumn{8}{l}{bias}\\
\hspace{1em}$opt$ & 0.072 & 0.051 & 0.151 & 0.146 & 0.016 & 0.229 & 0.239\\
\hspace{1em}$bal$ & 0.071 & 0.051 & 0.15 & 0.146 & 0.017 & 0.229 & 0.237\\
\hspace{1em}$ren$ & 0.089 & 0.025 & 0.15 & 0.226 & 0.01 & 0.23 & 0.237\\
\addlinespace[1ex]
\multicolumn{8}{l}{Variance}\\
\hspace{1em}$opt$ & 0.009 & 0.027 & 0.047 & 0.038 & 0.013 & 0.093 & 0.035\\
\hspace{1em}$bal$ & 0.02 & 0.059 & 0.095 & 0.084 & 0.031 & 0.2 & 0.081\\
\hspace{1em}$ren$ & 0.006 & 0.009 & 0.01 & 0.013 & 0.01 & 0.02 & 0.012\\
\bottomrule
\end{tabular}
\end{table}

\end{knitrout}

%
%
\clearpage
\section{Conclusion}

Statistical matching is set to become a valuable tool with the increasing amount of data created in this century. In this article, we propose new methods for matching two complex surveys. The proposed statistical matching methods are flexible depending on the type of analysis we want to perform. We can either have a one-to-one unit matching using stratified balanced sampling, or use the optimal output of the linear program, or finally use prediction using weighted averages.

Based on simulations, we observe that the proposed methods have lower cumulative mean squared error. A major assumption that persists in statistical matching is the conditional independence. Since in most cases this assumption is not satisfied, it is generally difficult to do anything other than assuming this assumption. Our method returns a mean squared error smaller in both cases where conditional independence assumption is satisfied or not. Thus, our results show that the proposed methods are less sensitive to a conditional independence defect. This suggests that they are more efficient and give a better quality statistical match.

%
%


\newpage


\begin{thebibliography}{}

	\bibitem[Alfons and Templ, 2013]{cran:laeken}
	Alfons, A. and Templ, M. (2013).
	\newblock Estimation of social exclusion indicators from complex surveys: The
	  {R} package {laeken}.
	\newblock {\em Journal of Statistical Software}, 54(15):1--25.
	
	\bibitem[Chen and Haziza, 2019]{chen2019recent}
	Chen, S. and Haziza, D. (2019).
	\newblock Recent developments in dealing with item non-response in surveys: a
	  critical review.
	\newblock {\em International Statistical Review}, 87:S192--S218.
	
	\bibitem[Deville and S\"arndal, 1992]{dev:sar:92}
	Deville, J.-C. and S\"arndal, C.-E. (1992).
	\newblock Calibration estimators in survey sampling.
	\newblock {\em Journal of the American Statistical Association}, 87:376--382.
	
	\bibitem[Deville et~al., 1993]{dev:sar:sau:93}
	Deville, J.-C., S\"arndal, C.-E., and Sautory, O. (1993).
	\newblock Generalized raking procedure in survey sampling.
	\newblock {\em Journal of the American Statistical Association}, 88:1013--1020.
	
	\bibitem[D'Orazio, 2019]{d2019statmatch}
	D'Orazio, M. (2019).
	\newblock {\em StatMatch: Statistical Matching or Data Fusion}.
	\newblock R Foundation for Statistical Computing, Vienna, Austria.
	\newblock R package version 1.4.0.
	
	\bibitem[D'Orazio et~al., 2006]{d2006statistical}
	D'Orazio, M., Di~Zio, M., and Scanu, M. (2006).
	\newblock {\em Statistical matching: Theory and practice}.
	\newblock John Wiley \& Sons, Hoboken (New Jersey).
	
	\bibitem[Dudoignon, 2018]{dud:18}
	Dudoignon, L. (2018).
	\newblock Fusion statistique de données d'enquêtes : dernières avancées
	  pour les mesures d'audience.
	\newblock Médiamétrie, 70 rue Rivay, 92532 Levallois.
	
	\bibitem[Guandalini and Till\'e, 2017]{guan:till:2017}
	Guandalini, A. and Till\'e, Y. (2017).
	\newblock Design-based estimators calibrated on estimated totals from multiple
	  surveys.
	\newblock {\em International Statistical Review}, 85:250--269.
	
	\bibitem[Hasler and Till\'e, 2014]{has:til:14}
	Hasler, C. and Till\'e, Y. (2014).
	\newblock Fast balanced sampling for highly stratified population.
	\newblock {\em Computational Statistics and Data Analysis}, 74:81--94.
	
	\bibitem[Haziza, 2009]{haz:09}
	Haziza, D. (2009).
	\newblock Imputation and inference in the presence of missing data.
	\newblock In Pfeffermann, D. and Rao, C.~R., editors, {\em Sample surveys:
	  Design, methods and applications}, pages 215--246, New York, Amsterdam.
	  Elsevier/North-Holland.
	
	\bibitem[Jauslin et~al., 2021a]{StratifiedSampling}
	Jauslin, R., Eustache, E., Panahbehagh, B., and Till\'e, Y. (2021a).
	\newblock {\em StratifiedSampling: Different Methods for Stratified Sampling}.
	\newblock R Foundation for Statistical Computing, Vienna, Austria.
	\newblock R package version 0.4.0.
	
	\bibitem[Jauslin et~al., 2021b]{jau:eus:til:21}
	Jauslin, R., Eustache, E., and Till\'e, Y. (2021b).
	\newblock Enhanced cube implementation for highly stratified population.
	\newblock {\em Japanese Journal of Statistics and Data Science}, 4:783--795.
	
	\bibitem[Kim et~al., 2016]{kim:ber:par:16}
	Kim, J.~K., Berg, E., and Park, T. (2016).
	\newblock Statistical matching using fractional imputation.
	\newblock {\em Survey Methodology}, 42(1):19--40.
	
	\bibitem[Kim and Tam, 2020]{kim2020data}
	Kim, J.-K. and Tam, S.-M. (2020).
	\newblock Data integration by combining big data and survey sample data for
	  finite population inference.
	\newblock {\em International Statistical Review}, \text{under press}:1--20.
	
	\bibitem[Panaretos and Zemel, 2020]{pan:zem:2020}
	Panaretos, V.~M. and Zemel, Y. (2020).
	\newblock {\em An Invitation to Statistics in Wasserstein Space}.
	\newblock Springer International Publishing.
	
	\bibitem[Renssen, 1998]{ren:98}
	Renssen, R.~H. (1998).
	\newblock Use of statistical matching techniques in calibation estimation.
	\newblock {\em Survey Methodology}, 24(2):171--183.
	
	\bibitem[Yang and Kim, 2020]{yang2020statistical}
	Yang, S. and Kim, J.~K. (2020).
	\newblock Statistical data integration in survey sampling: A review.
	\newblock {\em Japanese Journal of Statistics and Data Science}, 3:625--650.
	
	\end{thebibliography}

\end{document}